\documentclass[aps,pra,letterpaper,twocolumn,superscriptaddress]{revtex4-2}
\usepackage{amsmath,amssymb} 
\usepackage[T1]{fontenc}
\usepackage[utf8]{inputenc} 

\usepackage{ulem} 
\usepackage{soul}

\usepackage{graphicx} 

\usepackage{hyperref}
\hypersetup{colorlinks=true,citecolor={black},linkcolor={blue},urlcolor={blue}} 
\usepackage{xcolor}
\usepackage{bm}

\newcommand{\ve}{\varepsilon}

\begin{document}

\title{Trembling Motion of Exciton-Polaritons Close to the Rashba-Dresselhaus Regime}

\author{Wen Wen} 
\affiliation{Division of Physics and Applied Physics, School of Physical and Mathematical Sciences, Nanyang Technological University, 637371, Singapore}

\author{Jie Liang} 
\affiliation{Division of Physics and Applied Physics, School of Physical and Mathematical Sciences, Nanyang Technological University, 637371, Singapore}

\author{Huawen Xu} 
\affiliation{Division of Physics and Applied Physics, School of Physical and Mathematical Sciences, Nanyang Technological University, 637371, Singapore}

\author{Feng Jin} 
\affiliation{Division of Physics and Applied Physics, School of Physical and Mathematical Sciences, Nanyang Technological University, 637371, Singapore}

\author{Yuri G. Rubo} 
\email{ygr@ier.unam.mx}
\affiliation{Instituto de Energías Renovables, Universidad Nacional Autónoma de México, Temixco, Morelos, 62580, Mexico} 

\author{Timothy C. H. Liew} 
\email{timothyliew@ntu.edu.sg}
\affiliation{Division of Physics and Applied Physics, School of Physical and Mathematical Sciences, Nanyang Technological University, 637371, Singapore}

\author{Rui Su} 
\email{surui@ntu.edu.sg}
\affiliation{Division of Physics and Applied Physics, School of Physical and Mathematical Sciences, Nanyang Technological University, 637371, Singapore}
\affiliation{School of Electrical and Electronic Engineering, Nanyang Technological University, 639798, Singapore}

\begin{abstract}

We report the experimental emulation of trembling quantum motion, or \textit{Zitterbewegung}, of exciton polaritons in a perovskite microcavity at room temperature. By introducing liquid crystal molecules into the microcavity, we achieve spinor states with synthetic Rashba-Dresselhaus spin-orbit coupling and tunable energy splitting. Under a resonant excitation, the polariton fluid exhibits clear trembling motion perpendicular to its flowing direction, accompanied by a unique spin pattern resembling interlocked fingers. Furthermore, leveraging on the sizable tunability of energy gaps by external electrical voltages, we observe the continuous transition of polariton \textit{Zitterbewegung} from relativistic (small gaps) to non-relativistic (large gaps) regimes. Our findings pave the way for using exciton polaritons in the emulation of relativistic quantum physics.
\end{abstract}

\date{\today}

\maketitle

The quantum velocity operator depends on the spin of a particle in the presence of spin-orbital coupling. The particle in motion induces the precession of the spin, that in turn affects the particle velocity, leading to a non-straight mean trajectory. The resulting trembling motion, or \textit{Zitterbewegung} (ZB), was first discussed by Erwin Schrödinger for high-energy free electrons in 1930 \cite{schroedinger30}. For the Dirac equation, this intriguing but counterintuitive phenomenon arises as a consequence of combining special relativity and quantum mechanics, where positive- and negative-energy components of elementary spinor particles interfere. The experimental observation of such a phenomenon appears inaccessible for high-energy electrons because of the extremely small amplitude (about the Compton wavelength $\sim10^{-12}$ m) and ultrahigh frequency ($\sim10^{21}$ Hz) \cite{rusin09,gerritsma10}. In light of these challenges, different experimentally accessible systems have been theoretically proposed to emulate ZB and other relativistic quantum effects, in which parameter tunability serves as the key ingredient for complete emulations to reach different settings \cite{schliemann05,cserti06,lamata07,rusin07,vaishnav08,zhang08,longhi10,chen19,ye19,lavor21}. Certain experimental progress has been achieved in various physical systems, including trapped ions \cite{gerritsma10}, ultra-cold atoms \cite{leblanc13}, optical lattices \cite{dreisow10}, phononic crystals \cite{yu16}, and photonic microcavities \cite{lovett23}. However, they usually demand sophisticated setups at cryogenic temperatures or multiple fixed configurations for parameter tuning. Exploring a dynamically tunable platform at room temperature is highly desirable but remains challenging.

The trembling motion disappears in highly symmetric systems, and a notable example is the two-dimensional particle in the Rashba-Dresselhaus (RD) regime, which has been first discussed for electrons in quantum wells with the Rashba \cite{rashba59,bychkov84} and the Dresselhaus \cite{dresselhaus55} spin-orbital couplings of equal strength \cite{schliemann03,bernevig06}. The absence of ZB is a consequence of SU(2) symmetry of the Hamiltonian \cite{bernevig06}, and the fact that different components of velocity still commute in this case. More recently, the successful introduction of liquid crystal (LC) molecules  into microcavities has been shown to sustain synthetic spin-orbit coupling with broad tunability and realization of the RD Hamiltonian \cite{rechcinska19,krol21,li22,lempickamirek22}. Tuning is directly achieved by applying an electrical field to rotate the LC molecules, which allows to precisely perturb the RD Hamiltonian and provides a novel on-chip platform with dynamical tunability for the complete emulation of the ZB effect and further relativistic quantum effects at room temperature.


Perturbation of the RD Hamiltonian leads to the opening of a gap in the energy spectrum. Depending on the value of the gap, two regimes of the trembling motion can be realized: relativistic, where the gap is small and the ZB effect becomes clear, and non-relativistic, where the gap is large and ZB becomes non-distinguishable with high oscillation frequency and negligible amplitude. The
perturbation, in general, is equivalent to an artificial applied
magnetic field in the microcavity plane. The perturbed
RD Hamiltonian couples the motion in only one
direction to the spin precession, and the realization of ZB with liquid-crystal filled microcavities underscores
the potential of exciton polaritons as a solid-state
platform for emulations of relativistic quantum physics
in the 1+1 dimensions in the presence of an artificial
magnetic flux. Tunable and robust access to the relativistic
regime of the polariton dynamics could open new avenues
to study the peculiarities of many-body phenomena
in open-dissipative systems, in conditions when the
polariton-polariton interaction becomes important, and
the effects of the external potential and disorder, which
can be easily introduced in the microcavities.

In this study, we demonstrate the experimental emulation of trembling motion, or \textit{Zitterbewegung}, of exciton polaritons with electrical tunability at room temperature. By integrating birefringent LC into CsPbBr$_3$  perovskite microcavities, we achieve the synthetic RD Hamiltonian with broad tunability, which allows to continuously modulate the polariton band structures and the energy gap between two orthogonal spinor states. Under a continuous-wave resonant excitation at high momentum states, the polariton fluid exhibits an oscillatory trajectory and the spin precession in the real space, providing the unequivocal evidence of polariton ZB. By modulating the electrical voltages for tuning the RD Hamiltonian, we further observe the continuous transition of polariton ZB from relativistic dynamics to non-relativistic dynamics. The observation of trembling polariton motion in our work provides precise detection of the SU(2)-symmetry state, since small deviations from the RD Hamiltonian lead to significant deviations of the mean polariton trajectory from the straight path.

\begin{figure}[t]
	\includegraphics[width=0.48
    \textwidth]{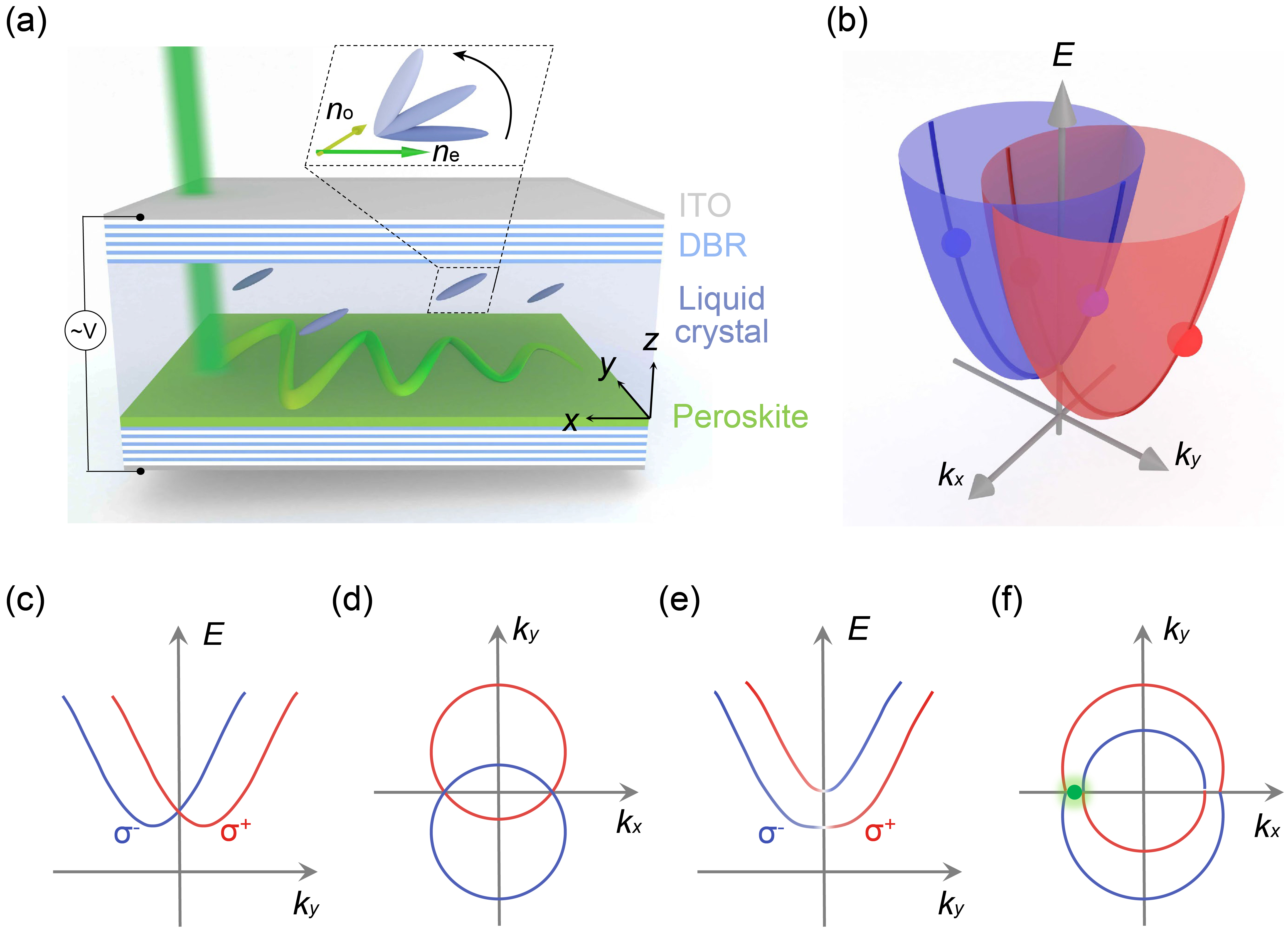} 
	\caption{
	Mechanism of exciton-polariton ZB effect with Rashba-Dresselhaus spin-orbit coupling at room temperature. 
	(a) Schematic illustration of polariton trembling motion in a perovskite microcavity with birefringent LC. 
	The rotation of LC molecules under applied voltages introduces synthetic spin-orbit coupling for continuously tuning 
	the energy gap between two spinor states, yielding the crossover from relativistic to non-relativistic dynamics 
	of polariton ZB effect. (b) Scheme of the exciton polariton band structure in the RD regime. 
	Dispersion (c) and iso-energy contour (d) of exciton polaritons in the exact RD regime. 
	Dispersion (e) and iso-energy contour (f) of exciton polaritons beyond the RD regime. 
	The green dot represents the resonant pumping point. 
	}
    \label{fig:1}
\end{figure}

 Fig.\ \ref{fig:1}(a) shows schematically the microcavity structure we employed in this work. Specifically, we embed single-crystalline CsPbBr$_3$ perovskites into highly reflective distributed Bragg reflectors (DBRs) made of TiO$_2$ and SiO$_2$. The single-crystalline CsPbBr$_3$, as demonstrated in our earlier studies, can sustain stable polariton fluids with long-range propagation and spontaneous coherence at room temperature \cite{su18,su20,feng21,su21}. We further fill the perovskite microcavity with birefringent LC and achieve synthetic spin-orbit coupling.  The LC employed here exhibit intrinsic birefringence characterized by an ordinary index $n_\mathrm{o}=1.53$ and an extraordinary index $n_\mathrm{e}=1.76$. As discussed in earlier studies, by tuning the external electric voltage applied to the microcavity, the LC molecules rotate, leading to tunable optical anisotropy, thus tunable energy gaps between orthogonally linearly polarized modes \cite{rechcinska19}. Under a proper voltage, an effective RD Hamiltonian 
\begin{equation}\label{RDHam0}
	H_0=\frac{\hbar^2}{2m}(k_x^2+k_y^2)-2{\alpha}k_y\sigma_z
\end{equation}
will arise, where $k_{x,y}$ are the components of the in-plane polariton wave vector, $m$ is the polariton effective mass, $\sigma_{x,y,z}$ are the Pauli matrices, and $\alpha$ is the RD coupling coefficient. This gives rise to spin-polarized bands split in momentum space [Fig.\ \ref{fig:1}(b)], when two orthogonally linearly polarized modes with opposite parity are brought into resonance. In the exact RD regime, as the polariton spins are split in momentum space, one would expect two intersecting spin circles for high momentum states at a given energy, where the polariton spins at the intersecting points are not well-defined [Fig.\ \ref{fig:1}(c, d)]. Therefore, these intersecting points at $k_y=0$ are two-fold degenerate with zero energy gap. Further tuning the electrical voltage, the LC molecules will continue to rotate and a tunable energy gap will arise between the spinor states in the whole momentum space [Fig.\ \ref{fig:1}(e, f)], which lays the foundation to completely simulate ZB with tunable dynamics.

Beyond the exact RD regime, the Hamiltonian with momentum-independent polarization splitting reads
\begin{equation}\label{RDPert}
	H=H_0-\frac{\ve}{2}(\sigma_x\cos\theta-\sigma_y\sin\theta),
\end{equation}
where $\ve$ denotes the energy gap between two linearly polarized states at $k=0$ and the angle $\theta$ defines their polarization orientation in the perovskite layer plane. 

In the Heisenberg picture, the velocities of exciton polaritons and the spin dynamics are described by 
\begin{equation}\label{HeisenDyn}
	\dot{x}=\frac{\hbar}{m}k_x, \quad \dot{y}=\frac{\hbar}{m}k_y-\frac{2\alpha}{\hbar}\sigma_z, \quad
	\dot{\bm{\sigma}}=\frac{1}{\hbar}[\bm{\sigma}\times\mathbf{b}],
\end{equation}
where $\mathbf{b}=(\ve\cos\theta,-\ve\sin\theta,4{\alpha}k_y)$ is the effective ``magnetic'' field acting on the polariton spin.

We consider a polariton wave packet of size $\sigma$ and with the X linear polarization, moving along the $x$-direction with the velocity $v_0={\hbar}k_0/m$. The mean displacement in the $x$-direction is not affected by the spin precession, ${\langle}x(t){\rangle}=v_0t$, while there appears a shift and trembling motion in the transverse $y$-direction (\cite{supplementary}). For very small gaps close to 0 ($\ve\ll2\alpha/\sigma$), and when the spectrum in the $y$-direction for wave vectors $\sigma^{-1}\ll2m\alpha/\hbar^2$ is  close to the Dirac spectrum of massless particles, only the shift is important. The packet is mainly displaced linearly with time:
\begin{equation}\label{LinShift}
	{\langle}y(t){\rangle}\simeq\sqrt{\frac{\pi}{2}}\,\sigma\sin\theta\,\frac{{\ve}t}{\hbar}.
\end{equation} 
The net shift of the polariton fluid saturates at about the size $\sigma$, and its position then exhibits oscillations of about the same amplitude, but with very long period about $h/\ve$. The presence of the shift allows precise detection of small values of the gap, restricted only by the polariton lifetime, which limits the observation time. The shift and ZB become the most pronounced at $\sigma\ve/2\alpha\sim1$. 
For large gaps, both the shift and the trembling amplitude decrease with the gap:  
\begin{equation}\label{Traject}
	{\langle}y(t){\rangle}\simeq\frac{2\alpha\sin\theta}{\ve}[1-\cos({\ve}t/\hbar)].
\end{equation}
Interestingly, the trembling motion in the $y$-direction is independent of the polariton fluid velocity $v_0$, and it takes place even when a static wave packet is excited.

\begin{figure}[t]
	\includegraphics[width=0.48
    \textwidth]{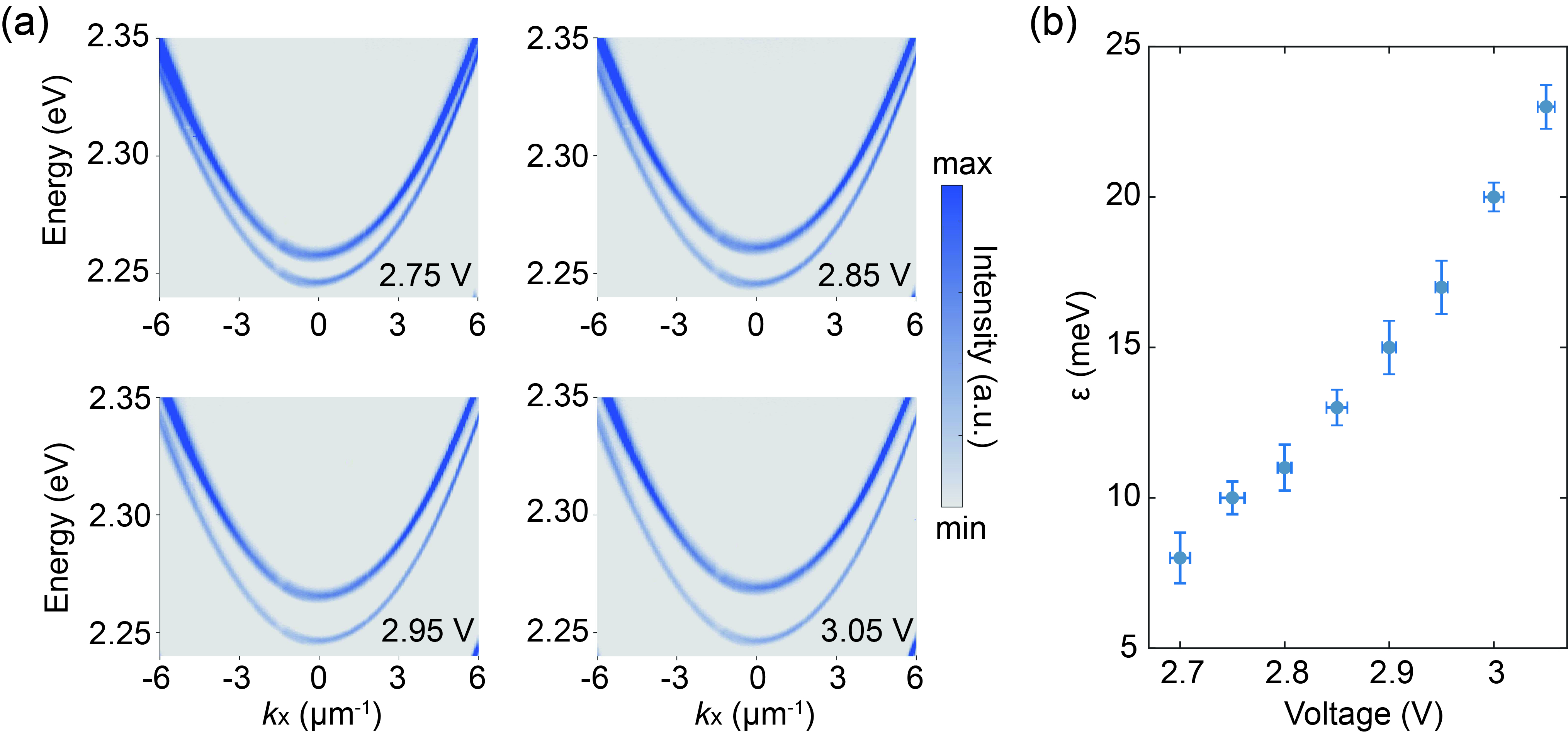} 
	\caption{Experimental exciton-polariton dispersions with tunable energy gaps as a function of applied voltage. 
	(a) Experimental angle-resolved PL spectra characterizing the polariton dispersion at $k_y=0$ cross section. 
	(b) Extracted energy gap of $\ve$ at $k=0$ as a function of applied voltage. 
    }
    \label{fig:2}
\end{figure}

In our experiments, we first confirm the high optical quality of the CsPbBr$_3$ microcavity by an optical microscopy image and the existence of long-range propagating polariton fluids (see Fig.\ S2 of the Supplemental Material). In order to fully simulate ZB, we demonstrate the continuous evolution of exciton-polariton band structure with tunable energy gap after introducing LC with controlled optical anisotropy. The perovskite microcavity is non-resonantly pumped by a continuous-wave laser at 2.713 eV and we collect the angle-resolved photoluminescence (PL) spectra to characterize the polariton dispersion under different applied voltages. At an applied voltage of $\sim2.60$ V, the polariton dispersion exhibits an unconventional splitting feature along $k_y$ and the $S_3$ Stokes parameter of the dispersion clearly illustrates the nature of spin-split bands with a high circular polarization degree of over 0.9, showing the occurrence of the RD regime (Fig.\ S3). Going beyond the RD regime at $\sim2.60$ V, we further increase the applied voltage and observe the gap opening $\ve$ at $k=0$. As displayed in Fig.\ \ref{fig:2}(a), the energy gap $\ve$ between the two spinor polariton states exhibits an increasing trend with growing of the applied voltage beyond 2.60 V. We further extract the energy gap $\ve$ as a function of the applied voltage [Fig.\ \ref{fig:2}(b)], showing a monotonic increase trend from $\sim8$ meV at 2.7 V to $\sim22.6$ meV at 3.05 V. This tunable energy gap between two spinor states in our system, serves as the key ingredient to simulate the trembling motion of exciton polaritons from the relativistic (small gaps) to the non-relativistic (large gaps) regimes.

\begin{figure}[t]
	\includegraphics[width=0.48
    \textwidth]{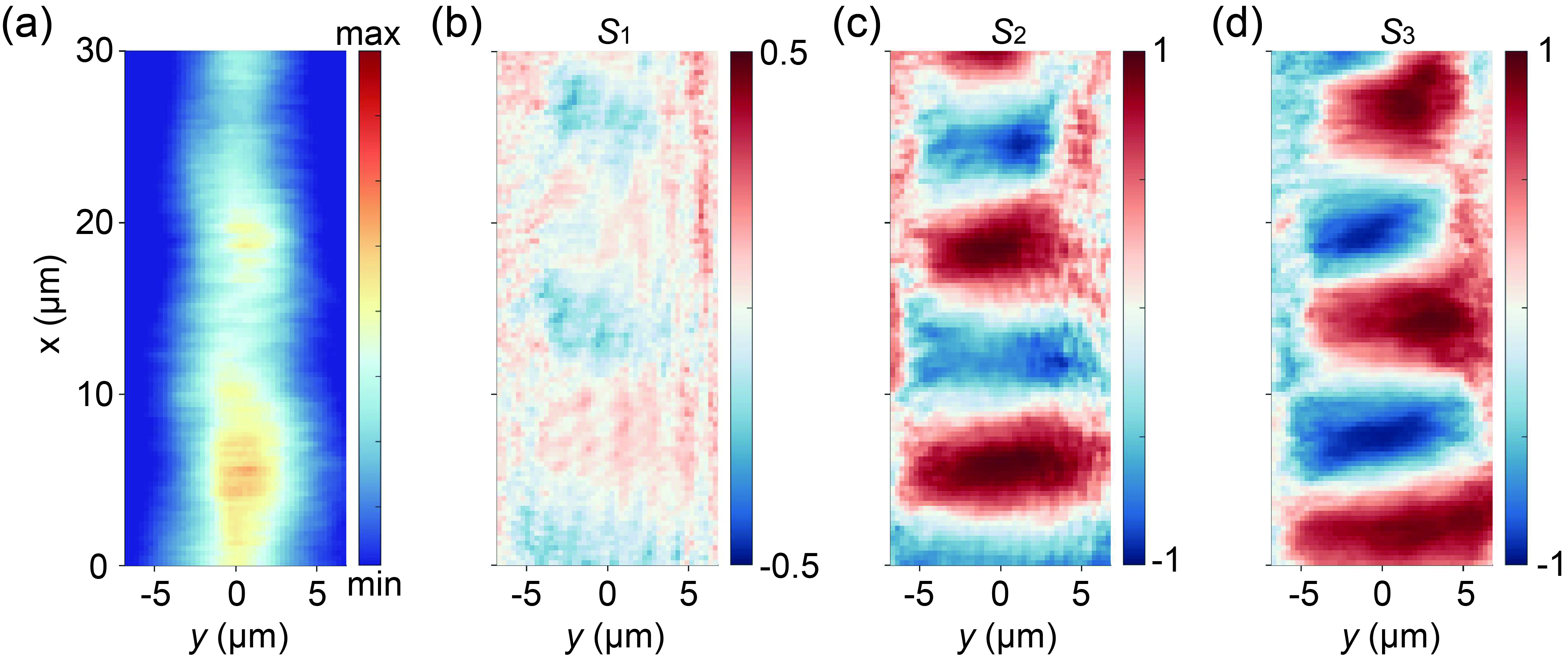} 
	\caption{Experimental exciton-polariton \textit{Zitterbewegung} in perovskite microcavity. 
	(a) Real-space image of exciton-polariton fluids at an applied voltage of 2.80 V. 
	(b-d) Polarization-resolved real-space image shown by spatial distribution of 
	$S_1$ (b), $S_2$ (c), and $S_3$ (d). 
    }
	
    \label{fig:3}
\end{figure}

\begin{figure*}[t]
	\includegraphics[width=0.65
	\textwidth]{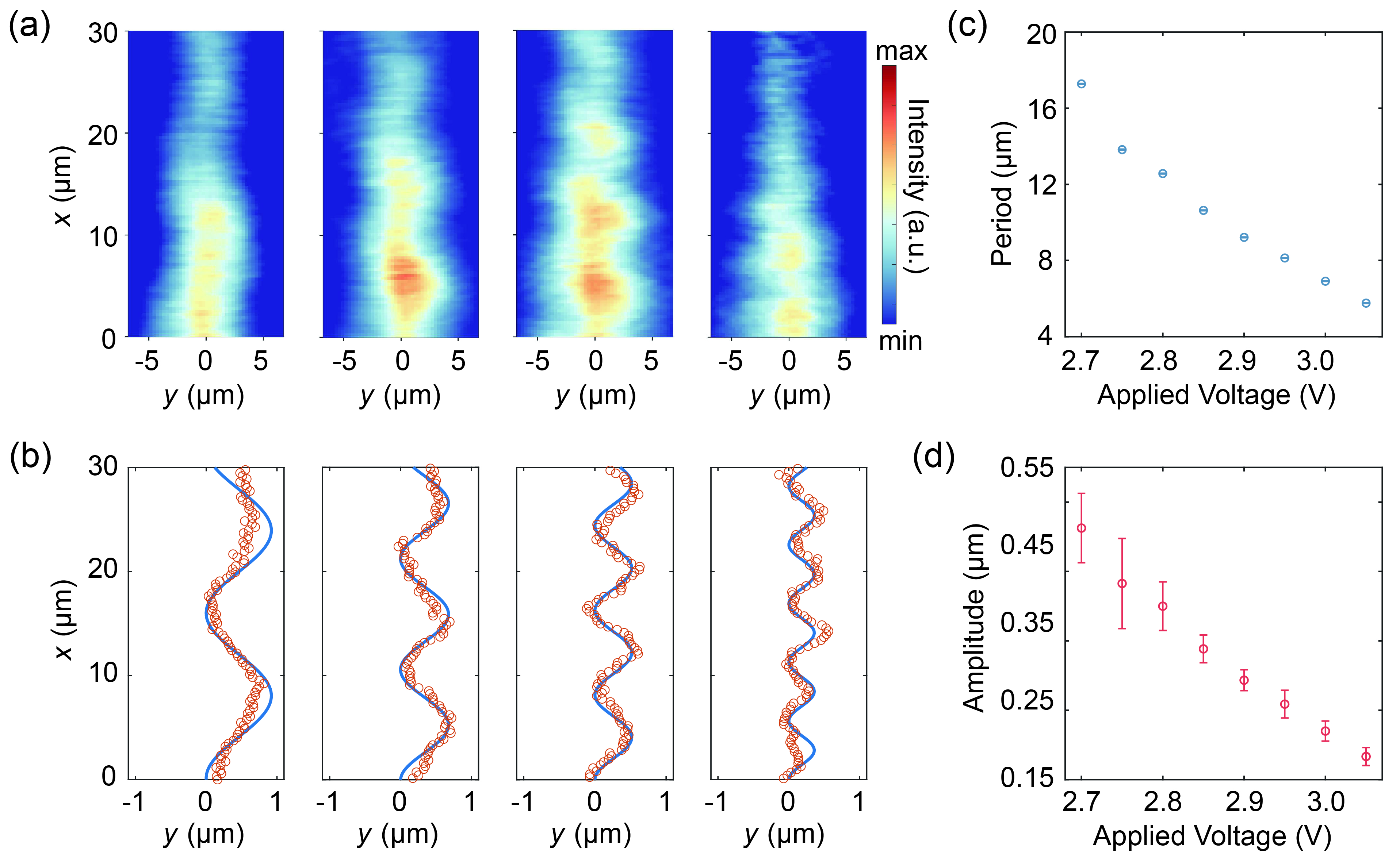} 
	\caption{
		Experimental demonstration of crossover from relativistic to non-relativistic regime in exciton-polariton \textit{Zitterbewegung}. 
		(a) Experimental real-space image of polariton trajectories 
		under applied voltages of 2.75, 2.85, 2.95, and 3.05 V. (b) The  extracted polariton center of mass (orange circle) from the experimental data in (a), compared with theoretically fitted polariton trajectories (blue curve). (c) and (d) Extracted 
		oscillation period and amplitude of exciton-polariton \textit{Zitterbewegung} under applied voltages 
		ranging from 2.7 to 3.05 V by fitting the experimental polariton trajectories.}
	
	\label{fig:4}
\end{figure*}

In light of the spinor polariton platform with a tunable energy gap, we investigate ZB by collecting the real-space trajectories of exciton-polariton fluids. Specifically, we employ a linearly polarized continuous-wave laser at 2.284 eV to resonantly pump the microcavity at $k_x=3.2$ µm$^{-1}$ in a transmission configuration (Fig.\ S4) and the spinor states exhibit typical energy splittings at this momentum (Fig.\ S5). As depicted in Fig.\ \ref{fig:3}(a), under a proper voltage of 2.80 V with an energy gap of 12 meV, the resonantly injected polariton fluid exhibits long-range propagation at room temperature, accompanied by a motion drift in the $y$-direction. Such non-trivial drift of the polariton fluid perpendicular to its propagation direction strongly evidences the existence of ZB. To gain further insight into the nature of ZB, we perform polarization-resolved real-space imaging of the polariton trajectories. The Stokes parameters of the polariton trajectories are calculated through 
$S_1=(I_H-I_V)/(I_H+I_V)$, $S_2=(I_D-I_{AD})/(I_D+I_{AD})$, and $S_3=(I_R-I_L)/(I_R+I_L)$, where $I_H$, $I_V$, $I_D$, $I_{AD}$, $I_R$, and $I_L$
denote polariton intensity under horizontal, vertical, diagonal, anti-diagonal linear polarizations, left- and right-handed circular polarizations, respectively. As shown in Fig.\ \ref{fig:3}(c,d), the spatial distribution of $S_2$ and $S_3$ resembles interlocked fingers for spinor components, indicating spin precession during the polariton propagation. This spin precession behavior directly correlates ZB to the interference between two spinor components, in line with previously predicted spin patterns in planar microcavities with TE-TM splitting \cite{sedov18} and our theoretical simulations (Fig.\ S6). It is noteworthy that such clear evidence of interference between spinor components is experimentally observed for the first time, thanks to the long-range polariton propagation and the tunable energy gap for sustaining a proper amplitude and frequency of ZB. This spin precession behavior is also observable at other applied voltages with different oscillation periods and amplitudes, serving as the solid proof of ZB effect (Fig.\ S7).

In order to unambiguously confirm the realization of the polariton ZB, we further demonstrate the tunable dynamics of ZB in different regimes. It can be conveniently achieved in our system where the energy gap can be continuously modulated with applied voltages. As shown in Fig.\ \ref{fig:4}(a) and Fig.\ S8, we collect the polariton trajectories under applied voltages ranging from 2.65 V to 3.05 V, where we experimentally observe distinct behaviors in terms of the periodicity and amplitude of the polariton oscillatory motion. Under a voltage of 2.65 V with a small energy gap $\ve\sim6$ meV, no clear oscillation feature can be observed as a result of long periodicity (Fig.\ S8), but the gradual shift of the wave packet is present. Increasing the voltage to 2.70 V with an energy gap of $\ve\sim8$ meV, clear oscillation features start to emerge, reaching into the relativistic regime. With the further increase of the voltage  and the energy gap, the oscillations become weaker and weaker with continuous decreases in period and amplitude and ZB becomes unobservable at 3.15 V (Fig.\ S8), suggesting the transition from relativistic to weak-relativistic and eventually to non-relativistic regimes. In order to quantify the evolution of period and amplitude with the applied voltage, we extract the center of mass trajectory \cite{sedov18} in their propagation by
\begin{equation}\label{CMyofx}
	{\langle}y(x){\rangle}=\frac{\int_{-\infty}^{\infty}yI(x,y)dy}{\int_{-\infty}^{\infty}I(x,y)dy},
\end{equation}
where $I(x,y)$ is the polariton intensity in real space.
Fig.\ \ref{fig:4}(b) and Fig.\ S8b present the extracted center of mass of polaritons under different applied voltages. We further fit these experimental trajectories with theoretical expressions to extract the amplitude and period of polariton ZB. Fig.\ \ref{fig:4}(c) and Fig.\ \ref{fig:4}(d) depict the amplitude and period of ZB  with error bars from the theoretical fittings as functions of the applied voltage, respectively. When the voltage increases from 2.7 V to 3.05 V, the corresponding period and amplitude decrease from 17.32 µm to 5.58 µm and from 0.51 µm to 0.19 µm, respectively.

In summary, we have unambiguously demonstrated the observation of exciton polariton \textit{Zitterbewegung} in a LC-filled perovskite microcavity at room temperature. Building on the unique Rashba-Dresselhaus Hamiltonian with sizeable tunability, one can conveniently modify the spinor band structures with tunable energy gaps by applied electric voltages. Under a resonant excitation, the injected polariton fluids exhibit long-range and ultrafast propagation with a counterintuitive trembling motion perpendicularly to its propagation direction. We also observe unique spin textures during the polariton propagation, resembling interlocked fingers for opposite spinor components which serves as the compelling evidence of interference between spinor states. Furthermore, leveraging on the sizable tunability in our platform, we continuously tune the energy gap between spinor states and realize continuous transition of polariton \textit{Zitterbewegung} from relativistic to non-relativistic regimes. Our work opens up exciting new possibilities to on-chip quantum emulators at room temperature for studying novel relativistic quantum effects, such as Klein’s paradox {\cite {lamata07}}. Given the interacting nature of spinor exciton polaritons, we also anticipate new possibility in investigating the interplay between relativistic quantum effects and strong nonlinearity, for instance, how the trembling motion or \textit{Zitterbewegung} will evolve in the presence of nonlinearity.

\textit{Acknowledgments.}
Wen Wen and Jie Liang contributed equally to this work.
R.S. and T.C.H.L. gratefully acknowledge funding support from the Singapore Ministry of Education via the AcRF Tier 2 grant (MOE-T2EP50222-0008) and the Tier 1 grant (RG80/23). 
R.S. also gratefully acknowledges funding support from Nanyang Technological University via a Nanyang Assistant Professorship start-up grant and the Singapore National Research Foundation via a Competitive Research Program (Grant No. NRF-CRP23-2019-0007). T. C. H. L. and H. X. gratefully acknowledge funding support from the Singapore Ministry of Education via the AcRF Tier 2 grant (MOE-T2EP50121-0020).
Y.G.R. and T.C.H.L. gratefully acknowledge funding support from PAPIIT-UNAM grant IN108524.

%

\end{document}